\documentclass[10pt,conference]{IEEEtran}

\usepackage{graphicx}
\usepackage{subfigure}
\usepackage{cite}
\usepackage{amssymb, amsmath}

\begin{document}

\bstctlcite{TurboRefShort:BSTcontrol}

\title{Pseudo-random Puncturing: A Technique to Lower the Error Floor of Turbo Codes}


\author{\authorblockN{Ioannis Chatzigeorgiou, Miguel R. D. Rodrigues, Ian J. Wassell}
\authorblockA{Digital~Technology~Group,~Computer~Laboratory\\
University~of~Cambridge,~United~Kingdom\\
Email:~\{ic231,~mrdr3,~ijw24\}@cam.ac.uk} \and
\authorblockN{Rolando Carrasco}
\authorblockA{School~of~EE\&C~Engineering\\
University~of~Newcastle,~United~Kingdom\\
Email:~r.carrasco@ncl.ac.uk}}

\maketitle
\begin{abstract}
It has been observed that particular rate-1/2 partially systematic
parallel concatenated convolutional codes (PCCCs) can achieve a
lower error floor than that of their rate-1/3 parent codes.
Nevertheless, good puncturing patterns can only be identified by
means of an exhaustive search, whilst convergence towards low bit
error probabilities can be problematic when the systematic output of
a rate-1/2 partially systematic PCCC is heavily punctured. In this
paper, we present and study a family of rate-1/2 partially
systematic PCCCs, which we call pseudo-randomly punctured codes. We
evaluate their bit error rate performance and we show that they
always yield a lower error floor than that of their rate-1/3 parent
codes. Furthermore, we compare analytic results to simulations and
we demonstrate that their performance converges towards the error
floor region, owning to the moderate puncturing of their systematic
output. Consequently, we propose pseudo-random puncturing as a means
of improving the bandwidth efficiency of a PCCC and simultaneously
lowering its error floor.
\end{abstract}

\IEEEpeerreviewmaketitle

\section{Introduction}
\label{Intro}

Although in certain applications, such as satellite communications,
link reliability is of essence and low rate codes are used to
support it, bandwidth occupancy is more important in wireless
communications and hence high rate codes are preferred. A high rate
convolutional code can be obtained by periodic elimination, known as
puncturing, of particular codeword bits from the output of a parent
low rate convolutional encoder. Extensive analyses on punctured
convolutional codes have shown that their performance is always
inferior to the performance of their low rate parent codes (e.g. see
\cite{Hagenauer88,Haccoun89}).

The performance of punctured parallel concatenated convolutional
codes (PCCCs), also known as punctured turbo codes, has also been
investigated. Design considerations have been derived by analytical
\cite{Acikel99,Babich02,Kousa02} as well as simulation-based
approaches \cite{FanMo99,Land00,Blazek02}, while upper bounds on the
bit error probability (BEP) were evaluated in
\cite{Kousa02,Chatzigeorgiou06b}. Punctured turbo codes are usually
classified as systematic, partially systematic or non-systematic
depending on whether all, some or none of their systematic bits are
transmitted \cite{Land00}. Recent papers
\cite{Land00,Blazek02,Chatzigeorgiou06b} have demonstrated that
partially systematic PCCCs yield lower error floors than systematic
PCCCs of the same rate.

In \cite{Chatzigeorgiou06c} we showed that rate-1/2 non-systematic
PCCCs can achieve error floors, which are lower even than those of
their rate-1/3 parent PCCCs. This interesting outcome is valid when
maximum-likelihood (ML) decoding is employed. When suboptimal
iterative decoding is used, the absence of received systematic bits
causes erroneous decisions, which prohibit the iterative decoder
from converging to the error floor. Nevertheless, we demonstrated
that rate-1/2 child codes, whose BEP performance converges towards
an error floor which is lower than that of their rate-1/3 parent
PCCC, can still be found by means of an exhaustive search. During
this process, the union bound on the BEP of each rate-1/2 punctured
PCCC is computed and compared to the union bound of the rate-1/3
parent PCCC. Note that the union bound coincides with the error
floor of the code for high values of $E_{b}/N_{0}$
\cite{Benedetto96a}. Punctured PCCCs that achieve a bound lower than
that of their rate-1/3 parent PCCC are selected.

Computation of the exact union bound on the BEP of a punctured PCCC
becomes intensive as the interleaver size increases. In
\cite{Chatzigeorgiou06d} we presented a simple technique to
approximate the union bound of a turbo code and we demonstrated that
this approximation is very accurate when a large interleaver size is
used. We used our technique to identify a family of rate-1/2
partially systematic PCCCs, which we called pseudo-randomly
punctured PCCCs (PRP-PCCCs). Although we did not explore their BEP
performance in detail, we observed that particular PRP-PCCC
configurations could achieve a lower error floor than that of their
parent codes.

This paper builds upon the work carried out in
\cite{Chatzigeorgiou06c} and \cite{Chatzigeorgiou06d}. Initially, we
provide analytical expressions for the parameters that influence the
bit error performance of PCCCs. We then evaluate those parameters
and compute the union bound approximations for both rate-1/3 parent
PCCCs and rate-1/2 PRP-PCCCs. We demonstrate that the latter always
exhibit a lower error floor than the former, when large interleaver
sizes are considered. In order to verify our theoretical analysis,
we compare analytic results to simulations for specific PCCC
configurations. The paper concludes with a summary of the main
contributions.

\section{Performance Evaluation of PCCCs}
\label{PerformanceEvaluation}

Turbo codes, in the form of symmetric rate-1/3 PCCCs, consist of two
identical rate-1/2 recursive systematic convolutional encoders
separated by an interleaver of size $N$ \cite{Berrou96}. The
information bits are input to the first constituent convolutional
encoder, while an interleaved version of the information bits are
input to the second convolutional encoder. The output of the turbo
encoder consists of the systematic bits of the first encoder, which
are identical to the information bits, the parity check bits of the
first encoder and the parity check bits of the second encoder.

The bit error probability $P_{b}$ of a PCCC employing ML soft
decoding, on an additive white Gaussian noise (AWGN) channel, is
upper bounded as follows
\begin{equation}
\label{PB} P_{b}\leq P_{b}^{\text{u}}
\end{equation}where the union bound $P_{b}^{\text{u}}$ is defined as
\begin{equation}
\label{PB_upper} P_{b}^{\text{u}}=\sum\limits_{w}P(w).
\end{equation}Here, the sum runs over all possible values of
input information weight $w$, with $P(w)$ being the contribution to
the union bound $P_{b}^{\text{u}}$ of only those codeword sequences
which were generated by input sequences of a specific information
weight $w$. An individual contribution $P(w)$ is given by
\cite{Benedetto96a,Ryan03}
\begin{equation}
\label{Pw} P(w)=\:\sum\limits_{d}\frac{w}{N}B_{w,d}
Q\left(\sqrt{\frac{2R\cdot E_{b}}{N_{0}}\cdot d} \right),
\end{equation}where $N$ is the interleaver size, $R$ is the code rate of
the turbo encoder and $B_{w,d}$ denotes the number of codeword
sequences having overall output weight $d$, which were generated by
input information sequences of weight $w$.

In \cite{Benedetto96a} it was shown that the union bound on the BEP
of a PCCC using a uniform interleaver of size $N$ coincides with the
average of the union bounds obtainable from the whole class of
deterministic interleavers of size $N$. For small values of $N$, the
union bound can be very loose compared with the actual performance
of turbo codes using specific deterministic interleavers. However,
for $N\!\geq\!1000$, it has been observed that randomly generated
interleavers generally perform better than deterministic interleaver
designs \cite{Hall98}. Consequently, the union bound provides a good
indication of the actual bit error rate performance of a PCCC
operating in the error floor region, when long interleavers are
considered.

Derivation of all coefficients $B_{w,d}$ becomes a computationally
intensive process as the interleaver size increases, especially when
punctured PCCCs are considered \cite{Chatzigeorgiou06d}. However,
the union bound can be approximated as follows
\begin{equation}
\label{PB_upper_approx} P_{b}^{\text{u}}\approx P(w\!=\!2),
\end{equation}when long interleavers are used. This approximation is based on a number of
observations:
\begin{enumerate}
\item Codeword sequences,
which were generated by input sequences having the minimum possible
information weight, become the main contributors to the bit error
rate performance, as the size $N$ of the interleaver increases
\cite{Benedetto96b,Chatzigeorgiou06d}.
\item Owning to the structure of the constituent encoders, the minimum
information weight of an input sequence is always equal to two
\cite{Benedetto96b}.
\end{enumerate}Therefore, $P(w\!=\!2)$ is the dominant contribution
to the union bound over a broad range of bit error probabilities
\cite{Chatzigeorgiou06d, Benedetto96b} and can be used to predict
the error floor of turbo codes.

Throughout this paper, we use the union bound approximation as the
basis for the analytic performance comparison of turbo codes. In
particular, if $\mathcal{P}$ and $\mathcal{P}'$ are two PCCCs using
long interleavers of identical size, we say that $\mathcal{P}$
yields a lower error floor than that of $\mathcal{P}'$ when their
bound approximations, $P^{\mathcal{P}}(2)$ and $P^{\mathcal{P}'}(2)$
respectively, satisfy
\begin{equation}
\label{P2_Condition} P^{\mathcal{P}}(2)<P^{\mathcal{P}'}(2).
\end{equation}The above condition can be expanded using (\ref{Pw}) as follows
\begin{equation}
\label{P2_Sum_Condition} \small{
\sum\limits_{d}B^{\mathcal{P}}_{2,d}
Q\left(\sqrt{\frac{2R^{\mathcal{P}}\: E_{b}}{N_{0}}\: d}
\right)<\sum\limits_{d}B^{\mathcal{P}'}_{2,d}
Q\left(\sqrt{\frac{2R^{\mathcal{P}'}\: E_{b}}{N_{0}}\: d} \right)}.
\end{equation}It was demonstrated in
\cite{Benedetto96b} that the free effective distance,
$d_{\text{f}}$, which conveys the minimum weight of a codeword
sequence for a weight-2 input information sequence, has a major
impact on the performance of a turbo code. Consequently, if
$d^{\mathcal{P}}_{\text{f}}$ and $d^{\mathcal{P}'}_{\text{f}}$
denote the free effective distances of $\mathcal{P}$ and
$\mathcal{P}'$ respectively, condition (\ref{P2_Sum_Condition})
collapses to
\begin{equation}
\label{Full_Condition} \small{B^{\mathcal{P}}_{2,d_{\text{f}}}
Q\left(\sqrt{\frac{2R^{\mathcal{P}}\: E_{b}}{N_{0}}\:
d^{\mathcal{P}}_{\text{f}}}
\right)<B^{\mathcal{P}'}_{2,d_{\text{f}}}
Q\left(\sqrt{\frac{2R^{\mathcal{P}'}\: E_{b}}{N_{0}}\:
d^{\mathcal{P}'}_{\text{f}}} \right)},
\end{equation}which only considers the first non-zero, that is the most
significant, term of each sum.

Function $Q(\xi)$ is a monotonically decreasing function of $\xi$,
where $\xi$ is a real number. Therefore, if $\xi_{1}$ and $\xi_{2}$
are real numbers, with $\xi_{1}>\xi_{2}$, we deduce that
$Q(\xi_{1})<Q(\xi_{2})$, and vice versa, i.e,
\begin{equation}
\label{Q_Condition} Q(\xi_{1})<Q(\xi_{2})\Leftrightarrow
\xi_{1}>\xi_{2}.
\end{equation}Consequently, inequality (\ref{Full_Condition})
reduces to
\begin{equation}
\label{Condition_Free_Effective_Condition_DiffRate}
R^{\mathcal{P}}d^{\mathcal{P}}_{\text{f}}>R^{\mathcal{P}'}d^{\mathcal{P}'}_{\text{f}},
\end{equation}if
\begin{equation}
\label{Condition_Coeff}B^{\mathcal{P}}_{2,d_{\text{f}}} \leq
B^{\mathcal{P}'}_{2,d_{\text{f}}}.
\end{equation}When the code rates are equal, the free effective
distance of turbo codes plays a role similar to that of the free
distance of convolutional codes, since the performance criterion
(\ref{Condition_Free_Effective_Condition_DiffRate}) is simplified to
\begin{equation}
\label{Free_Effective_Condition_SameRate}
d^{\mathcal{P}}_{\text{f}}>d^{\mathcal{P}'}_{\text{f}}.
\end{equation}

Expressions (\ref{Condition_Free_Effective_Condition_DiffRate}) and
(\ref{Condition_Coeff}) will be the basis for the comparison of the
BEP performance in the error floor region of two PCCCs.

\section{Determination of Parameters that Influence the Performance of
Turbo Codes}

We will now determine the various parameters that affect performance
for two classes of turbo codes: conventional rate-1/3 PCCCs and
pseudo-randomly punctured rate-1/2 PCCCs. The turbo codes considered
throughout this paper are symmetric, i.e., the two constituent
encoders are identical.

\subsection{Rate-1/3 PCCCs}

Criteria (\ref{Condition_Free_Effective_Condition_DiffRate}) and
(\ref{Condition_Coeff}) require knowledge of the free effective
distance $d_{\text{f}}$ and the coefficient $B_{2,d_{\text{f}}}$ of
each PCCC. In the remainder of the paper, we use the abbreviation
``Par'' to denote a rate-1/3 parent PCCC. Its free effective
distance $d^{\text{Par}}_{\text{f}}$ can be expressed as the sum of
the minimum weight $d_{\text{min}}$ of the codeword sequence
generated by the first constituent encoder, and the minimum weight
$z_{\text{min}}$ of the parity check sequence generated by the
second constituent encoder, when a sequence of information weight
$w\!=\!2$ in input to the PCCC
\begin{equation}
\label{Parent_Free_Eff_Generic}
d^{\text{Par}}_{\text{f}}=d_{\text{min}}+z_{\text{min}}.
\end{equation}Taking into account that the turbo codes are
symmetric and the weight $u_{\text{min}}$ of the systematic output
sequence is always 2 since $w\!=\!2$, we can write
\begin{equation}
\label{Parent_Free_Eff_Specific}
d^{\text{Par}}_{\text{f}}=(u_{\text{min}}+z_{\text{min}})+z_{\text{min}}=2+2z_{\text{min}}.
\end{equation}

The number $B^{\text{Par}}_{2,d_{\text{f}}}$ of codeword sequences,
generated by a turbo encoder using a uniform interleaver of size
$N$, can be associated with the number $B_{2,d_{\text{min}}}$ of
codeword sequences having weight $d_{\text{min}}$, generated by the
first constituent encoder, and the number $B_{2,z_{\text{min}}}$ of
parity check sequences having weight $z_{\text{min}}$, generated by
the second constituent encoder, if we elaborate on the expressions
described in \cite{Benedetto96a}. In particular, we obtain
\begin{equation}
\label{Parent_Turbo_ConstituentCodes}
B^{\text{Par}}_{2,d_{\text{f}}}=\frac{B_{2,d_{\text{min}}}\cdot
B_{2,z_{\text{min}}}}{\displaystyle \binom{N}{2}},
\end{equation}where $B_{2,d_{\text{min}}}$ and
$B_{2,z_{\text{min}}}$ return the same value, since they both
consider the same trellis paths. Note that the first index in the
above notations refers to the input information weight, which is
two.

It was shown in \cite{Benedetto96b} that good rate-1/3 PCCCs are
obtained when their feedback generator polynomial $G_{R}$ is chosen
to be primitive, whilst their feedforward generator polynomial
$G_{F}$ is different than $G_{R}$. The period $L$ of a primitive
polynomial is given by \cite{Macwilliams76}
\begin{equation}
\label{Period_L_ForPrimitive} L=2^{\nu}-1,
\end{equation}where $\nu$ is the order of the polynomial, or equivalently, the
memory size of each constituent code.

We demonstrated in \cite{Chatzigeorgiou06d} that when a primitive
feedback generator polynomial is used, the minimum weight
$z_{\text{min}}$ and the coefficient $B_{2,2,z_{\text{min}}}$ can be
expressed as
\begin{equation}
\label{Parent_FromPrevPaper}
\begin{split}
z_{\text{min}}&=2^{\nu-1}+2,\\
B_{2,d_{\text{min}}}&=B_{2,z_{\text{min}}}=N-L,
\end{split}
\end{equation}
respectively. Consequently, expression
(\ref{Parent_Free_Eff_Specific}) assumes the form
\begin{equation}
\label{Parent_Free_Eff_Primitive}
d^{\text{Par}}_{\text{f}}=6+2^{\nu},
\end{equation}whilst, if we combine (\ref{Parent_Turbo_ConstituentCodes}) and
(\ref{Parent_FromPrevPaper}), the coefficient
$B^{\text{Par}}_{2,d_{\text{f}}}$ can be expressed as a function of
the intlerleaver size $N$ and the period $L$, as follows
\begin{equation}
\label{Parent_Bwd_Primitive}
B^{\text{Par}}_{2,d_{\text{f}}}=\frac{2(N-L)^{2}}{N(N-1)}.
\end{equation}In the special case when the size $N$ of the
interleaver is an integer multiple of the period $L$ of the feedback
generator polynomial, i.e., $N\!=\!\mu L$, we can rewrite
(\ref{Parent_Bwd_Primitive}) as
\begin{equation}
\label{Parent_Bwd_MultipleOfL}
B^{\text{Par}}_{2,d_{\text{f}}}=\frac{2L(\mu-1)^{2}}{\mu(\mu L-1)}.
\end{equation}

\subsection{Rate-1/2 Pseudo-randomly Punctured PCCCs}

A high rate PCCC can be obtained by periodic elimination of specific
codeword bits from the output of a rate-1/3 parent PCCC. A
puncturing pattern $\mathbf{P}$ can be represented by a $3\times M$
matrix as follows:
\begin{equation}
\label{gen_punc_pattern} \mathbf{P}=\left[
\begin{matrix}
p_{1,1}&p_{1,2}&\ldots&p_{1,M}\\
p_{2,1}&p_{2,2}&\ldots&p_{2,M}\\
p_{3,1}&p_{3,2}&\ldots&p_{3,M}
\end{matrix}\right],
\end{equation}where $M$ is the puncturing period and $p_{i,m}\in\{0,1\}$, with $i\!=\!1,2,3$ and
$m\!=\!1,\ldots,M$. For $p_{i,m}\!=\!0$ the corresponding output bit
is punctured, otherwise it is transmitted. The first and second rows
of the pattern are used to puncture the systematic and parity check
outputs, respectively, of the first constituent encoder. The third
row determines which parity check bits from the output of the second
constituent encoder will be punctured.

Pseudo-random puncturing has been described in
\cite{Chatzigeorgiou06d}, in detail. It is applied to rate-1/3
PCCCs, which use primitive feedback generator polynomials, hence the
polynomial period $L$ is also given by
(\ref{Period_L_ForPrimitive}). The puncturing pattern can be
constructed once the parity check sequence
$\mathbf{y}\!=\!(y_{0},y_{1},\ldots,y_{L})$ for an input sequence
$\mathbf{x}\!=\!(1,0,\ldots,0)$ of length $L\!+\!1$, has been
obtained at the output of the first constituent encoder. As long as
a trail of zeros follows the first non-zero input bit, the component
encoder behaves like a pseudo-random generator, hence the parity
check bits from $y_{1}$ to $y_{L}$ form a pseudo-random sequence. We
set the elements of the second row of the puncturing pattern to be
equal to the bits of this pseudo-random sequence, but circularly
shifted rightwards by one, i.e., $p_{2,m+1}\!=\!y_{m}$ for
$m\!=\!1,\ldots,L$. Note that in pseudo-random puncturing, the
puncturing period $M$ is equal to the period $L$ of the feedback
polynomial, i.e., $M\!=\!L$. The first row of the pattern is set to
be the complement of the second row, thus $p_{1,m}\!=\!1-p_{2,m}$.
In order to achieve a code rate of $1/2$, we do not puncture the
parity check output of the second constituent encoder, hence all the
elements of the third row are set to one, i.e., $p_{3,m}\!=\!1$.

As an example, let us consider a rate-1/3 PCCC with generator
polynomials $(G_{F},G_{R})\!=\!(5,7)_{8}$ in octal form. The memory
size of each constituent encoder is $\nu\!=\!2$, thus the period of
$G_{R}$ is found to be $L\!=\!2^{2}\!-\!1\!=\!3$. Consequently, we
set the input sequence to $(1,0,0,0)$ and we obtain the parity check
sequence $(1,1,1,0)$ at the output of the first constituent encoder.
The block of the last $L\!=\!3$ parity check bits, i.e., $(1,1,0)$,
forms a pseudo-random sequence. If we circularly shift the bits of
this pseudo-random sequence to the right by one and map them to the
elements of the second row of the puncturing pattern, we obtain
$[0\;1\;1]$. Eventually the puncturing pattern, based on which the
rate-1/2 PRP-PCCC is generated from the rate-1/3 parent PCCC,
assumes the form
\begin{equation}
\label{example_punc_pattern} \mathbf{P}=\left[
\begin{matrix}
1&0&0\\
0&1&1\\
1&1&1
\end{matrix}\right].
\end{equation}We emphasize that the puncturing pattern depends on the
generator polynomials of the rate-1/3 parent PCCC, hence different
polynomials yield different puncturing patterns. Furthermore, a
rate-1/2 PRP-PCCC can be obtained only if the parent PCCC uses
primitive feedback generator polynomials.

We have previously determined \cite{Chatzigeorgiou06d} the minimum
weight $d'_{\text{min}}$ of the codeword sequence generated by the
first constituent encoder, when a sequence of information weight
$w\!=\!2$ in input to the rate-1/2 PRP-PCCC. In particular, we found
that
\begin{equation}
\label{Pseudo_dmin} d'_{\text{min}}=2^{\nu-2}+2.
\end{equation}The parity check sequence generated by the second constituent
encoder is not punctured, thus its minimum weight is also given by
(\ref{Parent_FromPrevPaper}). Therefore, we can compute the free
effective distance $d^{\text{PRP}}_{\text{f}}$ of a rate-1/2
PRP-PCCC as follows
\begin{equation}
\label{Pseudo_Free_Eff_Primitive}
\begin{split}
d^{\text{PRP}}_{\text{f}}&=d'_{\text{min}}+z_{\text{min}}\\
&=(2^{\nu-2}+2)+2^{\nu-1}+2\\
&=4+3(2^{\nu-2}).
\end{split}
\end{equation}

Every time a particular column $m$ of the puncturing pattern is
active during the $N$ time steps of the coding process, codeword
sequences having minimum weight $d'_{\text{min}}$ are generated.
Their exact number, $A_{m}$, can be computed using the expressions
in \cite{Chatzigeorgiou06d}. In particular, we find that for
$M\!=\!L$ the number of minimum-weight codeword sequences $A_{m}$,
generated when column $m$ is active, is given by
\begin{equation}
\label{Pseudo_Am_Generic} A_{m}=\left\{\begin{array}[\relax]{ll}
\left\lfloor N/M\right\rfloor-1,&\!\!\text{if}\:\left(N\:\text{mod}\:M\right)\!<\!m\\[3pt]
\left\lfloor N/M\right\rfloor,&\!\!\text{otherwise},
\end{array}\right.
\end{equation}where $(\xi_{1}\;\text{mod}\;\xi_{2})$ denotes the remainder of division
of $\xi_{1}$ by $\xi_{2}$, and $\lfloor\xi\rfloor$ denotes the
integer part of $\xi$. In order to facilitate our analysis, we
assume that the interleaver size $N$ is an integer multiple of the
puncturing period $M$, i.e., $N\!=\!\mu M$, where $\mu$ is a
positive integer. Hence, (\ref{Pseudo_Am_Generic}) collapses to
\begin{equation}
\label{Pseudo_Am} A_{m}=\mu-1,
\end{equation}since $(N\:\text{mod}\:M)$ is always zero and $m\!>\!0$.

It has been demonstrated in \cite{Chatzigeorgiou06d} that
minimum-weight codeword sequences can be obtained only when the
active column $m$ is in the range $2\leq m\leq M$; every time one of
these $M\!-\!1$ columns of the puncturing pattern is active, $A_{m}$
minimum-weight codeword sequences are generated. Consequently, the
total number of codeword sequences having weight $d'_{\text{min}}$
assumes the value
\begin{equation}
\label{Pseudo_Bdmin_Generic} B_{2,d'_{\text{min}}}=(M-1)A_{m},
\end{equation}or, equivalently
\begin{equation}
\label{Pseudo_Bdmin_Specific} B_{2,d'_{\text{min}}}=(L-1)(\mu-1),
\end{equation}where $M$ has been replaced by $L$, since they are
equal quantities and they can be used interchangeably.

Similarly to the second constituent encoder of the rate-1/3 parent
PCCC, the second constituent encoder of the rate-1/2 PRP-PCCC also
generates a total of $B_{2,z_{\text{min}}}$ sequences having weight
$z_{\text{min}}$, since its parity check output is not punctured.
Consequently, the coefficient $B^{\text{PRP}}_{2,d_{\text{f}}}$ or a
rate-1/2 PRP-PCCC can be expressed as
\begin{equation}
\label{Pseudo_Bwd_Primitive}
\begin{split}
B^{\text{PRP}}_{2,d_{\text{f}}}&=\frac{B_{2,d'_{\text{min}}}\cdot
B_{2,z_{\text{min}}}}{\displaystyle \binom{N}{2}}\\
&=\frac{\left[(L-1)(\mu-1)\right]\cdot(N-L)}{\displaystyle \binom{N}{2}}\\
&=\frac{2(L-1)(\mu-1)^{2}}{\mu(\mu L-1)},
\end{split}
\end{equation}invoking (\ref{Parent_Turbo_ConstituentCodes}), which
can be used when PCCCs employing uniform interleavers of size $N$
are considered.

\section{Performance Comparison of Analytic to Simulation Results}

Having evaluated the parameters that influence the performance of
the PCCCs under investigation, we are now in the position to explore
whether a rate-1/2 PRP-PCCC exhibits a lower bound approximation
than that of its rate-1/3 parent PCCC. We observe that
$d^{\text{PRP}}_{\text{f}}$ can be expressed in terms of
$d^{\text{Par}}_{\text{f}}$, if we subtract
(\ref{Parent_Free_Eff_Primitive}) from
(\ref{Pseudo_Free_Eff_Primitive})
\begin{equation}
\label{Pseudo_Parent_FreeEff}
d^{\text{PRP}}_{\text{f}}=d^{\text{Par}}_{\text{f}}-(2+2^{\nu-2}).
\end{equation}Coefficient
$B^{\text{PRP}}_{2,d_{\text{f}}}$ can also be represented in terms
of $B^{\text{Par}}_{2,d_{\text{f}}}$, if we divide
(\ref{Pseudo_Bwd_Primitive}) by (\ref{Parent_Bwd_MultipleOfL})
\begin{equation}
\label{Pseudo_Parent_Bwd}
B^{\text{PRP}}_{2,d_{\text{f}}}=\left(\frac{L-1}{L}\right)\:B^{\text{Par}}_{2,d_{\text{f}}}.
\end{equation}

According to (\ref{Condition_Free_Effective_Condition_DiffRate}) and
(\ref{Condition_Coeff}), if both conditions
\begin{equation}
\label{Condition1_Comparison}
\frac{1}{2}d^{\text{PRP}}_{\text{f}}>\frac{1}{3}d^{\text{Par}}_{\text{f}}
\end{equation}and
\begin{equation}
\label{Condition2_Comparison}
B^{\text{PRP}}_{2,d_{\text{f}}}<B^{\text{Par}}_{2,d_{\text{f}}}
\end{equation}are satisfied, a rate-1/2 PRP-PCCC yields a lower
bound approximation than that of its rate-1/3 parent code. We deduce
from (\ref{Pseudo_Parent_Bwd}) that
$B^{\text{PRP}}_{2,d_{\text{f}}}$ is always less than
$B^{\text{Par}}_{2,d_{\text{f}}}$, thus the second condition holds
true. The first condition assumes the following form, if we
substitute $d^{\text{PRP}}_{\text{f}}$ with its equivalent, based on
(\ref{Pseudo_Parent_FreeEff}),
\begin{equation}
\label{Full_Condition_PseudoVsPARENT_s4}
d^{\text{Par}}_{\text{f}}>6+3(2^{\nu-2}).
\end{equation}Nevertheless, we have shown in (\ref{Parent_Free_Eff_Primitive})
that the free effective distance of the parent PCCC is given by
$d^{\text{Par}}_{\text{f}}\!=\!6+2^{\nu}$, which can be rewritten as
$d^{\text{Par}}_{\text{f}}\!=\!6+4(2^{\nu-2})$. Therefore,
$d^{\text{Par}}_{\text{f}}$ is always greater than $6+3(2^{\nu-2})$,
and hence, both conditions are satisfied.

\begin{figure}[t]
    \centering
    \includegraphics[width=0.9\linewidth]{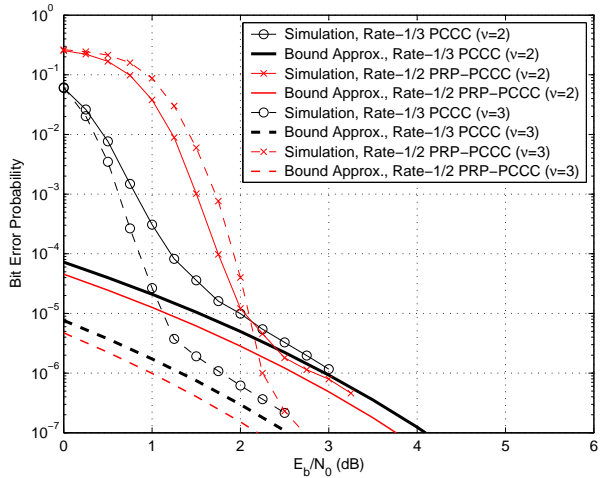}
    \caption{Comparison of bound approximations to simulation results. The exact log-MAP algorithm is applied over 8 iterations and an interleaver size of $1,000$ bits is used.}
    \label{Simulations}
\end{figure}

The outcome of this investigation reveals that rate-1/2 PRP-PCCCs
using long interleavers are always expected to yield a lower bound
approximation, or equivalently a lower error floor, than that of
their rate-1/3 parent codes.

Fig.\ref{Simulations} compares bound approximations to simulation
results for rate-1/3 parent PCCCs and rate-1/2 PRP-PCCCs of memory
size $\nu\!=\!2$ and $\nu\!=\!3$, over the AWGN channel. For
$\nu\!=\!2$, the generator polynomials of the PCCCs are taken to be
$(G_{F},G_{R})\!=\!(5,7)_{8}$, whilst for $\nu\!=\!3$, the PCCCs are
described by $(G_{F},G_{R})\!=\!(17,15)_{8}$. The component decoders
employ the conventional exact log-MAP algorithm \cite{Bahl74}. A
moderate interleaver size of $1,000$ bits has been chosen, so as to
allow the bit error rate performance of the PCCCs to approach the
corresponding bound approximations at BEPs in the region of
$10^{-6}$ to $10^{-7}$.

As expected, Fig.\ref{Simulations} confirms that for high values of
$E_{b}/N_{0}$, the BEP of each rate-1/2 PRP-PCCC is indeed lower
than that of the corresponding rate-1/3 parent code, whilst after 8
iterations the performance curves of all turbo codes approach the
respective bound approximation curves.

\section{Conclusion}
\label{conclusion}

In previous work \cite{Chatzigeorgiou06b,
Chatzigeorgiou06c,Chatzigeorgiou06d} we introduced techniques to
evaluate the performance of punctured PCCCs and we observed that, in
some cases, the error floor could be lowered by reducing the rate of
a PCCC from 1/3 to 1/2. Nevertheless, good puncturing patterns were
identified by means of an exhaustive search, whilst convergence
towards low bit error probabilities of those rate-1/2 PCCCs whose
systematic output was heavily punctured, had to be investigated.

In this paper, we established that rate-1/2 pseudo-randomly
punctured PCCCs, which form a subset of rate-1/2 partially
systematic PCCCs, not only approach the error floor region for an
increasing number of iterations but always yield a lower error floor
than that of their rate-1/3 parent codes. Consequently,
pseudo-random puncturing can be used to reduce the rate of a PCCC
from 1/3 to 1/2 and at the same time achieve a coding gain at low
bit error probabilities.

\bibliographystyle{IEEEtran}
\bibliography{IEEEabrv,TurboRefShort}

\end{document}